\documentclass[article, aps,superscriptaddress,floats,showpacs
]{revtex4}
\usepackage{cases}
\usepackage{graphicx}
\usepackage{amssymb}
\usepackage{ulem}

\usepackage{color}

\begin{document}

\title[Magnetic reconnection: from MHD to QED]{Magnetic reconnection: from
MHD to QED}
\author{S. V. Bulanov$^{1,2}$\\
$^1${\small {Kansai Photon Science Institute, 
National Institutes for Quantum and Radiological Science and Technology (QST), 
8-1-7 Umemidai, Kizugawa, Kyoto 619-0215, Japan}}\\
$^2${\small {A. M. Prokhorov Institute of General Physics, the Russian Academy of Sciences,\\
Vavilov street 38, 119991 Moscow, Russia}\\
e-mail:\,bulanov.sergei@qst.go.jp}\\
}
\begin{abstract}
The paper examines the prospects of using laser plasmas for studying novel
regimes of the magnetic field line reconnection and charged particle
acceleration. Basic features of plasma dynamics in the three-dimensional
configurations relevant to the formation of current sheets in a plasma  are addressed by analyzing exact self-similar solutions of the
magneto-hydrodynamics and electron magneto-hydrodynamics equations. Then the magnetic field annihilation in the ultrarelativistic limit is considered, when
the opposite polarity magnetic field is generated in collisionless plasma by
multiple laser pulses, in the regime with a dominant contribution
of the displacement current exciting a strong large-scale electric field.
This field leads to the conversion of the magnetic energy into the kinetic energy of
accelerated particles inside a thin current sheet. Charged particle
acceleration during magnetic field reconnection   is discussed when 
radiation friction and quantum electrodynamics effects become dominant.
\end{abstract}

\date{\today}
\maketitle


\section{Introduction}

One of the central problems of nowadays plasma physics, which has been studied for 70 years, 
is the reconnection of magnetic field lines.  The idea of magnetic field line reconnection stems
from the works aimed at finding mechanisms of charged particle acceleration in space plasmas \cite{Giovanelli}. 
 Then it evolved into the paradigm
embracing vast area of theories, experiments, and engineering problems related to
fundamental sciences and applications of magnetized plasmas \cite{BOOKSREVS}.
With the development of the high-power laser technology the magnetic reconnection in laser plasmas, foreseen a
number of years ago \cite{ABPP}, has recently attracted a great deal of attention of
several groups conducting  experiments and developing 
theory and computer simulations in this field  \cite{LASREC}. 

Magnetic reconnection, originally addressed within the framework of the
dissipative magneto-hydrodynamics (MHD), corresponds to the merging of oppositely directed magnetic fields
which leads to field reconfiguration, plasma heating,  jetting, and
acceleration, is one of the most complicated and important processes in
laboratory and space plasmas. It requires the violation of the condition of the 
magnetic field frozen-in. Laboratory
experiments in combination with analytical theory and sophisticated
numerical simulations reveal that reconnection occurs within a specific structure
of the current sheet formed on the site of critical points of the magnetic field \cite{SYROV, SYROV1, MHD-SIM}.

 The relativistic effects in magnetic reconnection important under the conditions of space and laser plasmas
manifest themselves in the displacement current effects playing a role of
``dissipation''  in the ultrarelativistic limit. They result in the strong electric field generation, 
which accelerates charged particles. Under the terrestrial
laboratory conditions, the relativistic regimes can be realized only with the
multi-petawatt power lasers. In the
limit of extremely high laser power, one should take into account the
effects of radiation friction  and of effects predicted by quantum
electrodynamics \cite{MTB, SVB-2015}. The radiation friction effects on the charged particle 
acceleration during the magnetic field line reconnection have recently been actively
discussed since they are related  to the interpretation of the high-energy
gamma-ray flares in astrophysics \cite{CRAB, CERUTTI}. 

Finally, the high power laser development
will provide the necessary  conditions for experimental physics where it
will become possible to study accelerated to ultrarelativistic energy
charged particles,  super high intensity electromagnetic waves and the
relativistic plasma dynamics. A fundamental property of the plasma  to
create nonlinear coherent structures will provide the conditions for
relativistic regimes of the magnetic field line reconnection,  making the
area of laser plasmas attractive for modeling the processes of key
importance for relativistic astrophysics \cite{LABASTRO}. In its turn, the laboratory astrophysics 
becomes one of the important motivations for the construction of the
ultra-high power lasers, let alone the importance of studies of astronomy, and fundamental
research in general.

{The text of this brief review article corresponds to the Hannes Alfven Prize lecture, presented by the author at the EPS-2016 conference on plasma physics. 
It is based on the author's results previously published (in Sections II--V) as well as it contains novel results (in Sections VI and VII) and the discussions of the prospects of their further developing.}

The paper discusses several theoretical problems related to nonlinear plasma
and charged particle dynamics near critical points of the magnetic field. In
the next Section we briefly describe typical magnetic field  patterns in the vicinities of critical points.
Then in Section III the reconnection 
within the framework of the MHD approximation is considered. We discuss a simple model 
of resistive reconnection, the exact self-similar
solutions of the MHD equations, which show how the current sheets can be
formed in the 3D geometry, and the Sweet-Parker-like current sheet parameters and its stability. 
Section IV addresses one of the basic mechanisms of the magnetic field line reconnection
in collisionless plasmas related to the fact that the curl of generalized
momentum is frozen into the electron component. We illustrate the magnetic field reconnection in collisionless plasmas with 
the Electron Magneto-Hydrodynamics (EMHD) 
by considering  a simple model 
of the reconnection due to the electron inertia effects and by presenting the exact self-similar
solutions of the EMHD equations, which show the piling up electron current. 
In Section V we discuss the relativistic limit using an example of the fast magnetic-field annihilation
 in the relativistic collisionless regime driven by two ultrashort high-intensity laser pulses, resulting in 
 the electric field generation and  charged particle acceleration. 
 In Section VI we analyze the regime when the radiation friction effects become significant limiting the achievable particle energy.
 Section VII is devoted to introduction of the quantum electrodynamics (QED) effects on the charged particle 
 motion in the vicinity of the magnetic null surface. In Conclusion, the main results are summarized.

\section{ Magnetic field patterns near critical points}

\label{S.PAT}

The magnetic field line reconnection in high conductivity plasmas occurs on the site of critical points of the field. 
Locally, in the vicinity of any arbitrary point, which we assume to be at the coordinates origin ${\bf x}=0$, we can expand the magnetic field as 
\begin{equation} 
 \label{eq:2.1}
B_i({\bf x},t)=B_i(0,t)+\partial_j B_i({\bf x},t)x_j+\partial_{jk}B_i({\bf x},t)x_jx_k... \,.
\end{equation}
We introduce  notations  $A_{ij}=\partial_j B_i|_{x_k=0}$, and $A_{ijk}=\partial_{jk} B_i|_{x_k=0}$ for the Jacobian and Hessian matrices of the magnetic field.
 Here and below $\partial_j=\partial x_j$ and summation over
repeated indexes is assumed. If the uniform part vanishes $B_i(0,t)=0$, a null
point of the magnetic field occurs at $x_i=0$, where 
\begin{equation}  \label{eq:2.2}
B_i=A_{ij}x_j+A_{ijk}x_jx_k+...\,.
\end{equation}

Locally, the topology of a magnetic field is determined
by the first nonzero term on the right-hand side
of expression (\ref{eq:2.2}).
If the matrix $A_{ij}$ vanishes, higher order terms in the right hand side of
equation (\ref{eq:2.2}) become dominant and we have an expression for the magnetic
field: 
$B_{i}=A_{ijk}x_{j}x_{k}.  \label{eq:3.strBfield}$
 
Let us assume that $A_{ij}$ is not equal to zero. 
It is well known that the equations, describing the behavior of the magnetic field lines in the field (\ref{eq:2.1}), have the form of the equations governing the behavior of a dynamical system: 
$dx_i/ds=A_{ij}x_j$  with $s$ being the parameter changing along the field line. The equilibrium position corresponds to the null point,
while the behavior of the trajectories (magnetic field lines) is determined
by the eigenvalues $\lambda_\alpha$ and eigenvectors $\mathbf{R}^{\alpha}$
of the matrix $A_{ij}$; ($\alpha=1,2,3$). By virtue of the condition $\mathrm{div}\,\mathbf{B}=0$,  
the trace of the matrix $A_{ij}$ is zero ($A_{kk}=0$) and
the sum of the eigenvalues vanishes, $\lambda_1+\lambda_2+\lambda_3=0$.
 
Below we mainly address magnetic configurations of two types. In
the first configuration one eigenvalue is equal to zero, let it be $\lambda_3=0$, 
while the other two are real numbers: $\lambda_{1,2}=\pm
\lambda^{\prime }$. In this case Eq.~(\ref{eq:2.1}) describes the
neighborhood of an X--line. This null line is a line of intersection of two
separatrix surfaces. In the second configuration all three eigenvalues are
non vanishing, and Eq.~(\ref{eq:2.1}) describes the neighborhood of a null
point which is a three-dimensional analog of the X--line. There is one
direction along which the magnetic field lines approach this point (or
leave it) and a separatrix surface along which magnetic the field lines
approach (or leave) its vicinity.
 
  A discussion of the plasma
dynamics in the vicinity of degenerate null points of the magnetic field, where $A_{ij}=0$, was
carried out in Ref.~\cite{STRUCT}. This problem is also connected with the studies  of 
the magnetic reconnection in collisionless plasmas within the framework of the Electron Magneto-Hydrodynamics (EMHD)
approximation \cite{EMHD, BPS-1992}.

\section{Reconnection in resistive magneto-hydrodynamics}
\subsection{System of MHD equations}

The paradigm of a magnetic reconnection can only be unambiguously formulated within the framework 
of the resistive magneto-hydrodynamics. Since the full system of dissipative MHD equations is cumbersome (e.g. see \cite{LLKIN, GIO-2016})
for the sake of brevity we use here the system of MHD equations where only the Ohmic dissipation is retained. 
It can be written in the form
\begin{equation}
\partial_t \rho +\nabla (\rho \mathbf{v})=0,
\label{eq.continuity}
\end{equation}
\begin{equation}
\partial_t \mathbf{v}+(\mathbf{v}\nabla )\mathbf{v}=-\frac{1}{\rho }\,\nabla p+\frac{1}{\rho }\,{\bf j}\times {\bf B} 
\label{eq.motion}
\end{equation}
for the plasma density $\rho$ and velocity ${\bf v}$. The plasma motion is caused 
by the pressure $p$ gradients and  Lorentz force ${\bf j}\times {\bf B}/c$.
 The relationship between the electric current density ${\bf j}$ and the magnetic field ${\bf B}$ is given by Maxwell equations:
 \begin{equation}
{\bf j}=\frac{c}{4 \pi}\nabla \times {\bf B},
\label{eq.Max1}
\end{equation}
\begin{equation}
\partial_t{\bf B}=-c \nabla \times {\bf E},
\label{eq.Max2}
\end{equation}
\begin{equation}
\nabla \cdot {\bf B}=0,
\label{eq.Max3}
\end{equation}
where, in Eq. (\ref{eq.Max1}), one neglects the displacement current. Ohm's law takes the form
\begin{equation}
{\bf E}=\frac{1}{c}\,{\bf v}\times {\bf B}+\frac{{\bf j}}{\sigma},
\label{eq.Ohm}
\end{equation}
with $\sigma$ being the plasma electric conductivity.

{In the ideal MHD limit, when the magnetic diffusivity $\nu_m=c^2/4\pi \sigma$ vanishes, the magnetic field lines 
cannot reconnect due to the Alfven's frozen-in theorem, which states 
that the magnetic flux trough the surface encircled by the contour moving with the plasma 
is conserved, i. e.  the magnetic field lines move along with the 
plasma.  In the limit when the electric conductivity tends to infinity $\sigma \to \infty$ the magnetic reconnection 
has been discussed in Refs. \cite{PUCCI, DELSARTO}.}

\subsection{ Kinematic Model of Reconnection in Collisional Plasmas}

The most simple example illustrating magnetic reconnection in resistive MHD is as follows. 

We consider 2D planar configuration with the magnetic field ${\bf B}(x,y,t)$ determined by 
the vector potential having the $z$ component $A(x,y,t)$: ${\bf B}=\nabla \times (A{\bf e}_z)$.
The plasma motion with the velocity field ${\bf v}(x,y)$ has a  stagnation point, where the 
velocity is the linear functions of the coordinates:
{
\begin{equation}
{\bf v}=W_{11}x\,{\bf e}_x+W_{22}y\,{\bf e}_y,
\end{equation}
with the velocity gradients $W_{11}$ and $W_{22}$ equal to  $\dot M_{11}/M^{-1}_{11}$  and $\dot M_{22}/M^{-1}_{22}$, respectively.
Here $M_{ij}$ with $i=1,2,\qquad j=1,2$ are the components of the deformation matrix, $\dot M_{ij}$ are their time derivatives, 
i. e. $M_{ij}(t)=\exp(\int W_{ij} dt)$, and ${\bf e}_x$ and ${\bf e}_y$ are unit vectors along the $x$ and $y$ axes.
}

Equations (\ref{eq.Max1}--\ref{eq.Ohm}) can be reduced to
\begin{equation}
dA/dt=\nu_m \Delta A,
\label{eq:Ares1}
\end{equation}
where $d/dt=\partial_t+({\bf v}\cdot\nabla)$. 
{For the magnetic field near the null-line of the form ${\bf B}=A_{12}y\,{\bf e}_x+A_{21}x\,{\bf e}_y$, } we look for the solution 
of Eq. (\ref{eq:Ares1}) in the form
\begin{equation}
A(x,y,t)=-\frac{1}{2}\left(A_{21}(t)x^2-A_{12}(t)y^2\right)+C(t),
\label{eq:Ares2}
\end{equation}
assuming the initial conditions are
\begin{equation}
A_{12}=A_{21}=A^{(0)}, \quad C(0)=0,
\end{equation}
which corresponds to the magnetic X-line.
The solution has the form
\begin{equation}
A(x,y,t)=-\frac{1}{2}A^{(0)}\left[\left(\frac{x}{M_{11}}\right)^2-\left(\frac{y}{M_{22}}\right)^2\right]+ \nu_m\int_0^t\left(\frac{M^2_{22}-M^2_{11}}{M^2_{11}M^2_{22}}\right)dt,
\label{eq:Arest}
\end{equation}

In the case of the time independent velocity gradients, $W_{11}=-W_{22}=w$ we have $M_{11}=\exp{(w t)}$ and   $M_{22}=\exp{(-w t)}$ Eq. (\ref{eq:Arest}) yields
\begin{equation}
A(x,y,t)=-\frac{1}{2}A^{(0)}\left[x^2\exp(-2 w t)-y^2\exp(2 w t)+ \frac{4\nu_m}{w}\sinh(2wt)\right].
\label{eq:Areswt}
\end{equation}

During the change of the magnetic field pattern the angle between the separatrices 
decreases. The instantaneous position of the separatrices is given by the expression
\begin{equation}
|y|=|x|\exp(-2wt).
\end{equation}
The magnetic field lines, which were at the position of the separatrices at $t=0$, move away. 
The evolution of the magnetic field line position is determined by
\begin{equation}
\exp(-2wt)x^2-\exp(2wt)y^2=(4 \nu_m/w)\sinh(2wt),
\label{eq:sep}
\end{equation}
which gives for the position of their intersection with the $x$-axis at $t\gg 1/w$
\begin{equation}
x(t)\approx \pm \left({4 \nu_m}/{w}\right)^{1/2}\exp(2wt).
\label{eq:sepx}
\end{equation}

The last terms in the r.h.s. of Eqs. (\ref{eq:Arest}) and  (\ref{eq:Areswt}) correspond to the last term in the r.h.s. 
of Eq. (\ref{eq.Ohm}), i.e. it is proportional to the time integral of the electric field occurring in the plasma due to 
the finite electric conductivity.

\subsection{Self-similar MHD motion near 3D critical points}

\label{S.SEL}

In the considered above kinematic model of resistive reconnection
 it was assumed that the magnetic field is weak and its effects on the plasma dynamics 
are negligibly small. In the limit when the nonlinearity
effects dominate,  we can use
self--similar solutions for the MHD equations in order to describe
the self-consistent evolution of the plasma
flow and of the magnetic field~\cite{BuOlsh1}.
 In these solutions,
that have the meaning of a local approximation,  the plasma
density $\rho(t)$ is uniform, and the plasma velocity field
and the magnetic field are given by
\begin{equation}
\label{eq:6.1}
v_i({\bf x},t)=W_{ij}(t)x_j,\qquad B_i({\bf x},t)=A_{ij}(t)x_j
\end{equation}
with the Jacobian matrices of the velocity field $W_{ij}$ and of the magnetic field $A_{ij}$, respectively. The dependence of the magnetic field on the
coordinates corresponds to a field which vanishes at ${\bf x}=0$. As  noted
in Section~\ref{S.PAT}, the nonuniform part of the magnetic field
 describes a neutral surface, a null line, or a null point, which is determined by the exact form of the matrix $A_{ij}$.

We introduce Lagrange coordinates which, for the
solutions under consideration, are related to  Euler's ones  by
\begin{equation}
\label{eq:6.2}
x_i=M_{ij}(t)x_j^0.
\end{equation}
The velocity Jacobian matrix $W_{ij}$ is expressed via the matrix $M_{ij}$ and its time derivative as $W_{ij}=\dot M_{ik}M_{kj}^{-1}$ with $M_{kj}^{-1}$ being the inverse matrix. 
In terms of the Lagrange variables, the solutions of the continuity equation (\ref{eq.continuity})
and of Faraday's equation Eq. (\ref{eq.Max2}) are
\begin{equation}
\label{eq:6.3}
\rho=\rho^{(0)}/D,\qquad A_{ij}=M_{ik}A_{kl}^{(0)}M_{lj}^{-1}/D.
\end{equation}
The superscript ``$^{(0)}$'' denotes initial values; $D(t)$ is the
determinant of the deformation matrix,
$M_{ij}~$: $D={\rm det}(M_{ij})$.

From the system of the MHD equations we obtain that
the matrix $M_{ij}$ obey
the following system of nonlinear ordinary differential equations:
\begin{equation}
\label{eq:6.3a}
\ddot M_{ij}=\frac{1}{4\pi \rho^{(0)}D}(M_{ik}A^{(0)}_{kl}A^{(0)}_{lj}-M_{sk}A^{(0)}_{kl}M_{lt}^{-1}M_{st}A^{(0)}_{tj}).
\end{equation}
In addition, it is  supposed that the plasma pressure
vanishes (see Ref.~\cite{BuOlsh1},
where the case $p\ne 0$ is considered, and Ref. \cite{BuSa} 
where the self-similar solution of the MHD equations describing the weakly ionized plasma have been obtained). 
The initial
conditions , at $t=0$, for  the deformation matrix are
$M_{ij}(0)=\delta_{ij},\, \dot M_{ij}(0)=W^{(0)}_{ij}$.

{
In the simplest three--dimensional magnetic configuration with a
null point, the magnetic field is
current-free and is given at the initial time by the diagonal
matrix 
\begin{equation}
A^{(0)}_{ij}={\rm diag}\,(a_{11},a_{22},-(a_{11}+a_{22})).
\label{eq:Adiag}
\end{equation}

If $a_{11}=a_{22}$ or $a_{11}=-a_{22}/2$ the magnetic field has azimuthal symmetry.
The case when $a_{11}=-a_{22}$ or $a_{11}=0$ for $a_{22}\ne 0$, or
$a_{22}=0$ for $a_{11}\ne 0$ the magnetic field given by Eq. (\ref{eq:6.1}) with $A^{(0)}_{ij}$ of the form (\ref{eq:Adiag}) corresponds to a quadrupole two--dimensional
configuration. 

The Eqs.~(\ref{eq:6.3a})
for the initial conditions of the form (\ref{eq:Adiag}) can be reduced to
\begin{equation}
\label{eq:6.5}
\left(
\begin{array}{ccc}
   \ddot M_{11} & \ddot M_{12} & 0 \cr \\
   \ddot M_{21} & \ddot M_{22} & 0 \cr \\
               0        & 0                     & \ddot M_{33}\cr 
            \end{array}
            \right)
            =\frac{(a_{22}-a_{11})(M_{11}M_{12}+M_{22}M_{21})}
{4\pi \rho^{(0)} M_{33}(M_{11}M_{22}-M_{12}M_{21})^2} 
\left(
\begin{array}{ccc} 
-a_{11} M_{21}&-a_{22} M_{22}&0\cr\\
               a_{11} M_{11}&-a_{22} M_{12}&0\cr\\
               0&0&0\cr
\end{array}
          \right).
\end{equation}

The solutions of these equations indicate the  formation of  current sheets.
These current sheets can be oriented at an arbitrary angle with
respect to the separatrix surface,
however the most probable configuration corresponds to
a current sheet  on the separatrix surface
directed along the minimum  gradient of the magnetic field.

Direct substitution into system (\ref{eq:6.5}) reveals that one can have a solution in which asymptotically, at 
$\tau=(t_0-t)\to 0$, all the components $M_{ij}$ remain finite except $M_{11}$ and $M_{12}$, which tend to zero as
\begin{equation}
\label{eq:M11tau}
M_{11}\approx \left(\frac{9 a_{11}^3 M_{21}^2(t_0)}{8\pi \rho^{(0)} M_{33}(t_0) M_{22} (t_0) (a_{11}-a_{22})} \right)^{1/3}\tau^{2/3}+...\, ,
\end{equation}
\begin{equation}
\label{eq:M21tau}
M_{21}\approx \left(\frac{9 a_{22}^3 M_{22}^2(t_0)}{8\pi \rho^{(0)} M_{33}(t_0) M_{21} (t_0) (a_{11}-a_{22})} \right)^{1/3}\tau^{2/3}+...\, ,
\end{equation}

According to Eq. (\ref{eq:6.3}), the magnetic
field gradients, $A_{12}\approx \tau^{-4/3}$,
as well as the velocity field gradients,
$W_{11}\approx \tau^{-1}$, and the plasma density, $\rho
\approx \tau^{-2/3}$, tend to infinity.
If both {\it a} and {\it b} have the same sign,
 the current sheet is orthogonal to the
separatrix surface, while if $ a_{11}$ and $ a_{22}$
have opposite signs  the current sheet is
parallel to the separatrix surface.

 In the generic case, by
expanding the solution near the
singularity, one can find that for $\tau \to 0$ 
the matrices $w_{ij}$ and $A_{ij}$ have the following forms
\begin{eqnarray}
\label{eq:6.8}
W_{ij}=
\left(
\begin{array}{ccc}
                     -2/3\tau& \tilde w_{12}& \tilde w_{13}\cr \\
              \tilde w_{21}/\tau^{2/3}&  \tilde w_{22}& \tilde w_{23}\cr \\
              \tilde w_{31}/\tau^{2/3}& \tilde w_{32}& \tilde w_{33}\cr 
             \end{array}
\right),~~~~~~
 A_{ij}=
\left(
\begin{array}{ccc}
                  \tilde a_{11}/\tau^{2/3}&  \tilde a_{12}& \tilde  a_{13} \cr \\
               \tilde   a_{21}/\tau^{4/3}& \tilde   a_{22}/\tau^{2/3}& \tilde  a_{23}/\tau^{2/3}\cr  \\
                \tilde  a_{31}/\tau^{4/3}&  \tilde a_{32}/\tau^{2/3}&  \tilde a_{33}/\tau^{2/3}\cr
                \end{array}
\right).
\end{eqnarray}
Here $\tilde w_{ij}$ and $\tilde a_{ij}$ are constant. }
These relationships
show that, as a result of the  development of the singular solution,
a quasi-one-dimensional magnetic configuration  is reached
during which the plasma is compressed toward a
surface and shear motion appears. This
implies that the magnetic collapse is
accompanied by a vortex collapse. 
The quasi-one-dimensional configuration corresponds to
 formation of a thin current sheet.
 
 {In the case of homogeneous electric conductivity the self-similar solutions describe the reconnection 
 of magnetic field lines with the rate which does not depend on the plasma resistivity. One may say that the reconnection 
 develops on ``ideal'' MHD time-scale (see also discussions of related problems in \cite{Terani-2015}).  
 According to Eqs. (\ref{eq:Arest}) and (\ref{eq:M11tau}) the arising electric field 
at $\tau \to 0$ is proportional to $\nu_m \tau^{-4/3}$, which, in particular, may lead to the burst of high energy particles. In the quasi-stationary configurations the rate of magnetic reconnection 
is substantially lower being determined by the plasma electric conductivity (see below).}

\subsection{Current Sheet}

In 2D magnetic configuration with a thin current sheet formed in a plasma on the site of the 
original X-line can be described by representing the magnetic field in terms of 
 a complex function $B(x,y)=B_{x}-iB_{y}$ of a
complex variable $\zeta =x+iy$. Introducing a complex potential $f(\zeta)=F(\zeta)-i A(\zeta)$ the 
complex magnetic field can be written as 
\begin{equation}
B_{x}-iB_{y}=d f/d \zeta.
\end{equation}

\begin{figure}[tbp]
\includegraphics[keepaspectratio=true,height=65mm]{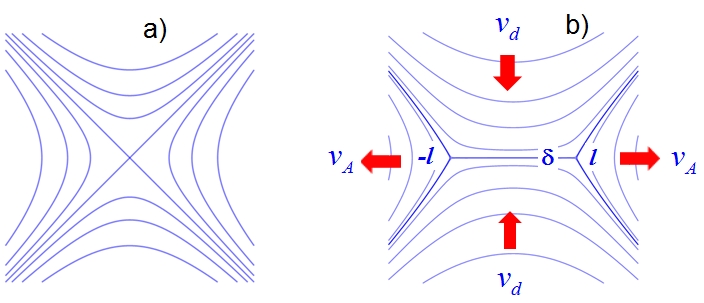}
\caption{Constant vector potential surfaces: $A={\rm constant}$, corresponding to magnetic field pattern in the vicinity of the X-line (a) and to the current
sheet of the thickness $\delta$ and width $2 l$ formed in the vicinity of the X-line. The red arrows show the plasma flow. 
The plasma flows in the current sheet with the velocity $v_d$ and flows out of the sheet with the 
velocity of the order of the Alfven velocity $v_A$ (b).}
\label{Fig2}
\end{figure}

In the case of the  X-line,  the magnetic field vanishes at the
coordinate origin. It is given by the complex potential $f(\zeta)=b \zeta ^{2}/2$. 
The magnetic field lines are hyperbolas
as we can see in Fig. \ref{Fig2} a. 
Under finite amplitude perturbations the magnetic X-line evolves to the magnetic
configurations of the form $B=b\sqrt{\zeta^2 -l^2}$, which describes the
magnetic field created by a thin current sheet between two points $\pm l$ \cite{SYROV1}. 
The magnetic field lines lie on the constant value surfaces of 
\begin{equation}
A(x,y)=\frac{b}{2}\mathrm{Re}\left\{ \zeta \sqrt{\zeta ^{2}-l^{2}}-\mathrm{Log}\left[ \zeta +\sqrt{\zeta ^{2}-l^{2}}\right] \right\} .
\end{equation}
They are shown in Fig. \ref{Fig2} b. The width of the current layer $l$ is
determined by the total electric current $J$ inside, and by the magnetic
field gradient, $b$. The current layer width is equal to
\begin{equation}
\label{eq:l-sheet}
l=\sqrt{4J/bc.}
\end{equation}

In the strongly nonlinear stage of the magnetic field and plasma evolution a
quite complex pattern in the MHD flow in the nonadiabatic region near the
critical point can be formed, with shock waves and current sheets as it follows from the results of the dissipative magnetohydrodynamics
simulations of the current sheet formation near the X-line \cite{MHD-SIM}.

\subsection{Sweet-Parker-like Model of Current Sheet}

The Sweet-Parker model of the current sheet \cite{SWPRK} is one of the central and  well known results
in the theory of magnetic field line reconnection. According to this model the plasma with the magnetic field frozen in it 
flows in the current sheet with the velocity equal to $v_d$. Due to a finite resistivity the magnetic field line reconnect inside the 
current sheet of a half-thickness $\delta=\nu_m/v_d$ (see Fig. \ref{Fig2}). The magnetic field line tension and the gradient of the plasma pressure lead to 
the plasma ejection along the current sheet with the velocity of the order of the Alfven velocity $v_A=B/\sqrt{4\pi \rho}$. From the 
continuity condition it follows that $\rho_pv_d l=\rho_s v_A \delta $, where $\rho_p$ and $\rho_s$ is the plasma density outside and inside 
the current sheet, respectively. Assuming for the sake of simplicity that  $\rho_p= \rho_s=\rho$, we can find a relationship between the reconnection 
velocity $v_d$ and the width of the current sheet. It reads $v_d=(\nu_m v_A/l )^{1/2}$. Now we take into account that the magnetic field at the ends of the 
current sheet is determined by the gradient of the field $b=|A_{ij}^{(0)}|$ at the magnetic field $X$-line at the site of which the current sheet is formed and 
by the width of the current sheet $l$, i.e. $B=bl$. The current sheet width depends on the electric current $J$ carried by the sheet according to Eq. (\ref{eq:l-sheet}). 
As a result we find the out-flow 
velocity. It is equal to $v_A=\Omega_A l$, with $\Omega_A=b/\sqrt{4\pi \rho}$. This gives for the reconnection velocity:
\begin{equation}
v_d=\sqrt{\Omega_A \nu_m}.
\end{equation} 
The current sheet thickness is equal to $\delta=(l \nu_m/v_A)^{1/2}=(\nu_m/\Omega_A)^{1/2}$. {In other words, the reconnection rate, $v_d/v_A$ is equal to $S^{-1}$}.
It is determined by the magnetic Reinolds number (it is also called the Lundquist number), calculated for the current sheet thickness as $S={v_A}/{v_d}=l/\delta$) as  $v_d=v_A/S$. 
In the case under consideration, the Lundquist number {calculated for the current sheet width,  $l$,} can be expressed 
in the form $S_l=v_A l/\nu_m=4 J/c\nu_m \sqrt{4\pi \rho}$.

{We note that in the standard Sweet-Parker model, in contrast with the analyzed above the case of the current sheet formation near the magnetic X-line, the width of the current sheet $l$ and the magnetic field $B$, 
i.e. also the Alfven velocity $v_A$, are independent parameters, which has important implications for the current sheet stability (see below).

\subsection{Effect of  plasma flow along the current sheet on the tearing mode instability }

Tearing mode instability, resulting in the current sheet break up into filaments \cite{FKR, FKR0}, has been intensively studied 
for a number of years in connection with various applications  \cite{BOOKSREVS}. According to the results of the theory,
 the static equilibrium with the current sheet, e. g. described by the Harris solution \cite{HARRIS}, for which the equilibrium magnetic field is equal to ${\bf B}=B^{(0)}\tanh(y/\delta){\bf e}_x$, is always 
 unstable. At the same time, laboratory experiments \cite{SYROV}, computer simulations \cite{MHD-SIM} and the interpretation 
 of observations in space show that the current sheets may exist for a relatively long time. On the other hand side, 
 there are computer simulations showing the development of the tearing mode and the secondary magnetic reconnection 
 via formation of the chains of the magnetic islands \cite{SECMAGISL} (see also  Refs. \cite{PUCCI, DELSARTO, UZD}).  
 The secondary magnetic islands has been {\it in situ} observed in an ion diffusion region of the Earth magnetosphere \cite{INSITU-2010}.

 The controversy can partially be resolved if one takes into account the effects of an inhomogeneous plasma flow along the current sheet. 
 The plasma flows-in the current sheet with the velocity $v_d$ and flows out with the velocity depending on the coordinate as $v=w x$  (see Fig. \ref{Fig2}).
 As shown in Ref. \cite{TeaFlow}, the tearing mode can be stabilized if its growth rate is less than the velocity gradient. 
 Since the growth rate depends on the wavenumber of the mode this condition determines the stability wavelength range, i.e. the width  of stable (unstable) current sheet. 

For example, let us consider the regime of resistive tearing mode, for which the growth rate is given by (see Refs. \cite{FKR0})
%
\begin{eqnarray}
\label{eq:gFKR}
\gamma_{\small{FKR}}\,\tau_A\approx 
\left\{
\begin{array}{ccc}
                    S^{-3/5}(k \delta)^{-2/5}(1-k^2\delta^2)& \qquad k \delta S^{1/4}\gg 1\cr \\
          S^{-1/2}& \qquad k \delta S^{1/4}\approx 1\cr \\
               S^{-1/3}(k \delta)^{2/3} & \qquad k \delta S^{1/4}\ll 1\cr 
             \end{array}
\right.,~~~~~~
\end{eqnarray}
with $\tau_A=\delta/v_A$.

Taking into account the inhomogeneous along the current sheet plasma motion with the velocity $v=w x$ we can find that 
the tearing mode is stabilized provided $w>\gamma_{\small{FKR}}$. Since, according to Eq. (\ref{eq:gFKR}) the instability growth rate depends on the 
wavenumber, this condition determines the stability domain in the parameter space.

In the short wavelength limit, $k \delta S^{1/4}\gg 1$,
 the tearing instability threshold wavenumber is equal to  \cite{TeaFlow}
\begin{equation}
\label{eq:k1}
k_1\approx\frac{S}{\sqrt{2}\delta (w \delta^2/\nu_m)^{5/2}}.
\end{equation} 
The tearing mode is stabilized for $k>k_1$.

In the long wavelength regime, when $ k \delta S^{1/4}\ll 1$, the tearing mode is stabilized  for
\begin{equation}
\label{eq:k2}
k<k_2\approx\left(\frac{w}{v_A}\right)^{3/2}\sqrt{S\delta}.
\end{equation} 

If the velocity gradient is above the maximum growth rate, i. e. $w>\tau_A^{-1}S^{-1/2}$, the current sheet is fully stabilized otherwise it is unstable  
in the wavenumber range: $k_2<k<k_1$. 

In the case, when the magnetic reconnection occurs as a result of perturbing the Harris-like equilibrium, usually the current sheet length $l$, the Alfven velocity $v_A$
and the reconnection velocity $v_d$ are independent parameters being determined by the initial conditions and by the amplitude and scale-length of the perturbations 
imposed from the boundary. 

Using the instability threshold (\ref{eq:k1}) and taking into account that the growth rate is maximal for the perturbation wavelength $2\pi/k_1$  
of the order of the current sheet length and estimating the velocity gradient as $w=v_A/l$ we obtain that the current sheet of the length 
\begin{equation}
l<\delta S^{3/7}
\end{equation} 
 is stable \cite{TeaFlow}. The stabilization  condition around the growth rate maximum,  $w>\tau_A^{-1}S^{-1/2}$, is equivalent to the constraint $l<\delta S^{1/2}$. 
 In the long wavelength limit the stability condition is $l>2\pi/k_2$ with $k_2$ given by Eq. (\ref{eq:k2}), which is equivalent to $l<\delta S$.

As we have seen above, in the quasi-stationary configuration described within the framework of the Sweet-Parker model of magnetic reconnection the current sheet 
width, its thickness,  and  the Lundquist number are related to each other as $l=\delta S$. This implies that for $S\gg 1$ the current sheet being under the conditions of the marginal stability for the  perturbation wavelength  
equal to the current sheet length is unstable in the short wavelength limit with a relatively narrow wavenumber range near $k\delta \leq 1$ .  The development of the tearing mode and the secondary magnetic reconnection 
 via formation of the chains of the magnetic islands \cite{SECMAGISL, PUCCI, DELSARTO, UZD, COMISSO-2016} can also be interpreted as a formation of secondary current sheets near secondary 
 X-lines formed as the instability result according to the scenario presented in Ref. \cite{BSS-1977}.

}

\subsection{ EMHD-equations}
In collisionless plasmas, instead of the magnetic field to be frozen-in, the curl of generalized momentum, 
\begin{equation}
{\bf p}_{\alpha}=m_{\alpha} {\bf v}_{\alpha}+(e_{\alpha}/c){\bf A}, 
 \label{eq:gen-momentum}
\end{equation}
is frozen-in for each of the plasma components $\alpha$, i.e.
\begin{equation}
\partial_t \nabla \times {\bf p}_{\alpha}=\nabla  \times \left({\bf v}_{\alpha}\times\nabla\times {\bf p}_{\alpha}\right), 
 \label{eq:alphaMHD}
\end{equation}
where ${\bf v}_{\alpha}$ is the corresponding flow velocity, in which the generalized vorticity $\nabla \times {\bf p}_{\alpha}$ is frozen.

The Electron Magneto-hydrodynamics (EMHD) considers the dynamics of the electrons only. 
The ions are assumed to be at rest and the quasi-neutrality is fulfilled, i.e. the electron and ion densities are equal to each other, $n_e=n_i$. In this case, the Hall effect is dominant, i.e. the electron inertia determines the
relationship between the electric field and the electric current density
carried by the electron component, the magnetic field evolution is described
by the equation (see \cite{EMHD, BPS-1992}) 
\begin{equation}
\partial _{t}(\mathbf{B}-\Delta \mathbf{B})=\nabla \times \left[ \left(
\nabla \times \mathbf{B}\right) \times (\mathbf{B}-\Delta \mathbf{B})\right], 
 \label{eq:EMHD}
\end{equation}
which corresponds to the condition of electron generalized vorticity, 
$\mathbf{\Omega }=\mathbf{B}-\Delta \mathbf{B}$, to be frozen into the electron component
moving with the velocity $\mathbf{v}_{e}=c\nabla \times \mathbf{B/}4\pi
n_{0}e$, which follows from Eq. (\ref{eq.Max1}). Here, in Eq. (\ref{eq:EMHD}),
 the spatial scale is chosen to be equal to the collisionless
electron skin-depth, $d_{e}=c/\omega _{pe}$, and the time is measured in units of $\omega
_{Be}^{-1}=m_{e}c/eB$. The range of frequencies described by the EMHD
equations is given by $\omega _{Bi}<\omega <\omega _{Be}$.

In the linear approximation Eq.~(\ref{eq:EMHD}) describes the propagation of
whistler waves, propagating in a plasma with the magnetic field $\mathbf{B_{0}}$.
 For the whistler waves the relationship between the wave frequency $\omega$ and
the wave vector ${\bf k}$, is 
$\omega =|\mathbf{k}|(\mathbf{k}\cdot \mathbf{B_{0}})/(1+|\mathbf{k}|^{2})$. 
It follows from this relationship that in a weakly
inhomogeneous magnetic field the critical points are the points and lines
where $|\mathbf{B_{0}}|=0$ or/and $(\mathbf{k}\cdot \mathbf{B_{0}})=0$.

\subsection{Basic Mechanism of Collisionless Magnetic Reconnection}

To illustrate the basic mechanism of the magnetic field line reconnection in collisionless plasmas
we consider a simple 2D model similar to the model used above in the case of resistive 
MHD reconnection. In 2D geometry the three component magnetic field can be expressed in terms of the 
$z$-components of the vector potential $A_{||}$ and magnetic field $B_{||}$ as
\begin{equation}
{\bf B}(x,y,t)=\nabla \times (A_{||}(x,y,t){\bf e}_z)+B_{||}(x,y,t){\bf e}_z.
 \label{eq:BAzBz}
\end{equation}
The EMHD equations can be written as 
\begin{equation}
\frac{d}{dt}(A_{||}-\Delta A_{||})=0
 \label{eq:emhdA}
\end{equation}
and
\begin{equation}
\frac{d}{dt}(B_{||}-\Delta B_{||})+\{B_{||},\Delta B_{||}\}=\{A_{||},\Delta A_{||}\},
 \label{eq:emhdB}
\end{equation}
where $d/dt=\partial_t+({\bf v}\cdot\nabla)$ and $\{g,f\}=\partial_x g \partial_y f-\partial_x f \partial_y g$ are 
 Poisson brackets for functions $f$ and $g$.

We seek the solution of Eqs. (\ref{eq:emhdA},\ref{eq:emhdB}) in the form
\begin{equation}
A_{||}=A_{21}x^2+A_{12}y^2+C(t) \quad {\rm and} \quad B_{||}=w(t) x y,
\label{eq:Acolles}
\end{equation}
assuming the initial conditions are
\begin{equation}
A_{21}=-A_{12}=A^{(0)}, \quad C(0)=0,  \quad w(0)=w^{(0)}.
\end{equation}
The electron velocity in the $(x,y)$-plane for the quadruple magnetic field $B_{||}=w x y$ is equal to 
\begin{equation}
{\bf v}_{\perp}=\nabla\times(B_{||}{\bf e}_z) =w (x {\bf e}_x-y {\bf e}_y).
\end{equation}
The solution of Eqs. (\ref{eq:emhdA},\ref{eq:emhdB}) has the form
\begin{equation}
w(t)=w^{(0)} \quad {\rm and} \quad
A(x,y,t)=A^{(0)}\left[\exp(-2w^{(0)}t)x^2-\exp(2w^{(0)}t)y^2-4 \sinh (2w^{(0)}t)\right].
\label{eq:BAcollest}
\end{equation}
Similarly to the the case of resistive MHD reconnection we can see from expression (\ref{eq:BAcollest}) that 
the magnetic field lines move with respect to the magnetic separatrices, i.e. the magnetic field is not frozen-in the electron component. 
This is a simplest example of the magnetic field line reconnection in collisionless plasmas due to the 
electron inertia effects.

The electron inertia effects make the reversed magnetic field configuration
unstable against tearing modes~\cite{FKR}, which result in magnetic
field line reconnection. The slab equilibrium configuration with a magnetic
field given by $\mathbf{B_{0}}=B^{(0)}_{z}\mathbf{e}_{z}+B^{(0)}_{x}(y/\delta)\mathbf{e}_{x}$, 
where $B^{(0)}_{x}(y/\delta)=B^{(0)}\tanh(y/\delta)$ is the function that gives the current sheet magnetic
field, is unstable with respect to perturbations of the form $f(y)\exp
(\gamma t+ikx)$ with $k\delta<1$. For this configuration one has $(\mathbf{k}\cdot \mathbf{B_{0}})=0$ 
at the surface $y=0$. The growth rate of the
tearing mode instability is \cite{BPS-1992, BASOVA} $\gamma \approx
(1-k\delta)^{2}\Delta ^{\prime 2}/k\delta^{2}$.

In Ref. \cite{AVI} (see also \cite{FCal-2001}) in Fig. 3 the results of a numerical solution of Eq.~(\ref{eq:EMHD}) 
in a 2D geometry with magnetic field 
$\mathbf{B}(x,y,t)=\left(\nabla \times A_{||}\right) \times \mathbf{e_{\bot }}+B_{||}\mathbf{e_{\Vert }}$ 
are shown. The unperturbed configuration is chosen to be a current sheet,
infinite in the $x$-direction, that separates two regions with opposite
magnetic field. Both the line pattern of generalized vorticity, $\Omega
=A_{||}-\Delta A_{||}$, and of the magnetic field show the formation of
quasi--one--dimensional singular distributions in the electric current
density and in the distribution of the generalized vorticity. During this process the magnetic
field topology changes.

{It is worth noticing that electron magnetohydrodynamics has been used for studying the electromagnetic 
filametation instability, magnetic island and vortex structure formation, and ion acceleration in the electric current carrying plasmas \cite{FCal-1997, SAKAI-2002}. }

\subsection{Nonlinear Pile-up of Magnetic field near Magnetic Null-Points in
EMHD}

Now we discuss the regime of the nonlinear accumulation of  the magnetic
field energy near the critical points. This regime is described in terms of a self-similar solutions of the
EMHD equations~\cite{BPS-1992}. It is well known that the formal solution of Eq.~(\ref{eq.Max2})
is given by the formula obtained by Cauchy:
\begin{equation}
\label{eq:8.3}
 B_i({\bf x},t)-\Delta B_i({\bf x},t)=D_e\left(\frac{\partial x_i}{\partial x^0_j}\right)(B_j({\bf
x^0},0)-\Delta B_j({\bf x^0},0)).
\end{equation}
Here $D_e\equiv {\rm Det}\left({\partial x_k^0}/
{\partial x^0_l}\right)$ is the Jacobian of the
transformation from the Lagrange variables $x^0_i$ to the Euler coordinates $x_j$.
In regimes typical for the EMHD
approximation ions are at rest and,
due to  plasma quasineutrality, the electron motion is
incompressible and $D_e =1$. The Euler and the Lagrange
variables are related to
each other by the formula
$x_i=x^0_i+\xi_i({\bf x^0},t)$, where $\xi_i({\bf x^0},t)$
is the displacement of the electron fluid element from its
initial position $x^0_i$. From
 Maxwell equations, $4\pi n e {\bf v}/c={\rm curl}\,{\bf B}$, taking into account the condition $n={\rm
const}$ and ${\bf v}={\partial \xi/\partial t}$, we
obtain that the function $\xi({\bf x^0},t)$
obeys the equation
\begin{equation}
\label{eq:8.4}
 \frac{\partial \xi_i}{\partial t} =
-\varepsilon_{ijk}\left(\frac{\partial x_j^o}{\partial
x^0_l}\right) \left(\frac{\partial B_l({\bf x^0},t)}{\partial x^0_k}\right),
\end{equation}
where $\varepsilon_{ijk}$ is the antisymmetric Ricci tensor.

In the self-similar solutions the magnetic field spatial
and time dependences are given by
\begin{equation}
\label{eq:8.5}
B_i({\bf x},t)=A_{ijk}(t)x_jx_k.
\end{equation}
This expression describes the magnetic field pattern in
the vicinity of a null point of the third
order. In the two-dimensional case it is the
line of intersection of three separatrix surfaces (in
the frame of the standard MHD model
magnetic field line reconnection in the vicinity of such
lines has been investigated in Ref.~\cite{STRUCT}.)
From Eq.~(\ref{eq:8.4}) it follows that the
 fluid velocity of the electron component is a linear function of
 coordinates: ${\partial \xi_i/\partial
t}=w_{ij}(t)x_j$. Taking into account the
relationships given by Eq.~(\ref{eq:6.2}) we obtain
\begin{equation}
\label{eq:8.6}
A_{ikl}(t)=M_{ij}(t)A_{jmn}^{(0)}M_{mk}^{-1}(t)M_{nl}^{-1}(t),
\end{equation}
while the deformation matrix $M_{mk}(t)$ obeys the equation
\begin{equation}
\label{eq:8.7}
\dot M_{ij}=-2\varepsilon_{ikl}M_{lm}M_{nk}^{-1}A_{mnj}^{(0)}.
\end{equation}

In the case of 2D magnetic configuration, the exact solutions of the EMHD equations 
given by Eqs. (\ref{eq:BAcollest}) correspond to the self-similar plasma motion with the magnetic field of the form 
$B_i({\bf x},t)=A_{ij}(t)x_j+A_{ijk}(t)x_jx_k$. We note that in general case this is the magnetic configuration 
(\ref{eq:2.2}), which for $A_{ij}=0$ is  structurally unstable.

\section{ Relativistic Regime of Magnetic Field Annihilation}

\subsection{Relativistic Tearing Mode Instability of a Thin Current Sheet}

The study of the magnetic field reconnection, which was started in \cite{Giovanelli}, initially was aimed at the explanation of
the generation of suprathermal particles during solar flares and substorms
in the earth's magnetosphere. 
It is well known, the acceleration of charged particles during the
magnetic reconnection is due to the electric field generated by the fast
change of the magnetic field. This electric field is considered to be of an
inductive nature. The change of the magnetic field is caused by
the redistribution of the electric current in plasmas. The electric
current configuration may change due to the development of the tearing mode
instability, which leads to the electric current filamentation and, in the
strongly nonlinear regime, breaks up the current sheet into separated pieces. 
The generated electric field magnitude is of the order of $E\approx (v/c)B$, 
where $v$ is the typical value of the plasma velocity.

The current sheet can be unstable against the so-called tearing mode, which 
leads to the electric current filamentation and change of the magnetic field 
topology \cite{FKR}. In the relativistic plasmas the tearing mode has been studied in 
Refs. \cite{ZK}. In the limit of high anisotropy when the current velocity 
of plasma electrons, $u_0=j/en$, is substantially larger than the thermal velocity, the tearing mode 
growth rate is given by \cite{SAKAI-2002}
\begin{equation}
\gamma_{TM}=\frac{k u_0}{1+\gamma_{0}^3|k|c^2/2\pi n_0 \delta Z_i e^2}\sqrt{\frac{m_e}{m_i}\gamma_{0}^3}
\end{equation}
Here $k$ is the perturbation wavenumber, $n_0$ and $\delta$ are the electron density and the current sheet thickness, 
and {the electron gamma-factor $\gamma_{0}=1/\sqrt{1-u_0^2/c^2}$, which shows the electron inertia effects}. 

\subsection{Inductive Electric Field Generation by Ultra Intense Two Laser Pulses in
Underdense Plasmas}

As we have seen, magnetic reconnection is accompanied by a
current sheet formation, where the oppositely directed
magnetic fields annihilate. The magnetic-field annihilation
has been investigated within the framework of dissipative
magnetohydrodynamics (see  literature cited
in Ref. \cite{YuGU}). In ultrarelativistic plasma, it becomes principally
different because the electron current density has the upper limit \cite{SYROV2} $j_{lim}=enc$.
Due to the relativistic constraint on the particle velocity (which
never exceeds the speed of light in vacuum) the electric current
can sustain only a limiting magnetic-field strength. {In other words, in ultrarelativistic limit the Ohm low 
should me modified (see also discussions in Refs. \cite{FP-2015, FP-2016} ). 
In addition, in relativistic plasmas the problem of magnetic field connection is quite nontrivial (also see Refs. \cite{Asenjo-2015, FP-2016}).}

The development of high power lasers allows accessing
new regimes of magnetic-field annihilation. When a high intensity
laser pulse interacts with a plasma target the accelerated
electron bunches generate strong regular magnetic fields.
Computer simulation results of two co-propagating laser pulse interaction 
with underdense laser plasmas has been presented in Refs. \cite{PING, YuGU}.
 In Refs. \cite{YuGU}, a fast magnetic-field annihilation in relativistic
collisionless plasma driven by two collinear ultraintense
femtosecond laser pulses is studied with particle-in-cell (PIC)
simulations. { Since in the ultrarelativistic regime the electric current
can sustain only a limiting magnetic-field strength the displacement current cannot be neglected. 
The displacement current causes strong electric field with the amplitude of the order of that of the magnetic field.  
This has been
demonstrated in Ref. \cite{YuGU}, where it has also been shown  that the induced electric field accelerates charged particles within
the current sheet.}

\section{Charged Particle Acceleration}

\subsection{Electric and Magnetic Field Configuration and Charged Particle Motion in the Non-adiabatic Region}

{ A fully developed tearing mode, either it is the primary or secondary mode,  results in a current sheet modulations  with the 
space scale equal to $2s$. The distance $s$ can also be considered as a size of magnetic islands.
Local structure of the magnetic field configuration can be analytically described as (see Ref. \cite{BSS-1977} and Fig. 1 therein) 
}
\begin{equation}
B_x(x,y)-i B_y(x,y)=b\mathrm{Re}\left\{ \zeta \sqrt{\frac{\zeta ^{2}-l^{2}}{s^2-\zeta ^{2}}}\right\}.
\label{eq:AxyDCS}
\end{equation}

Due to the magnetic
field line tension the plasma is thrown out. The local topology of the magnetic field lines 
within the break up describing this configuration is given by the function of a complex variable 
$B(\zeta )=B_{0}\zeta /\sqrt{s^{2}-\zeta^{2}}$. The magnetic field
lines lie on the surfaces of constant vector potential, 
\begin{equation}
A(x,y,t)=\mathrm{Re}\left\{ B_{0}\sqrt{s^{2}(t)-\zeta ^{2}}\right\} .
\label{eq:BreakUP-B}
\end{equation}
Due to the dependence of the function $s$ on time, an electric field parallel
to the $z$ axis arises. It is given by \cite{SYROV1}
\begin{equation}
E(x,y,t)=-\frac{1}{c}\partial _{t}A
=-\frac{1}{c}\frac{B_{0}s(t)\dot{s}(t)}{\sqrt{s^{2}(t)-\zeta ^{2}}}.
\label{eq:BreakUP-B1}
\end{equation}

In the vicinity of the null line we have a quadrupole structure of the
magnetic field $B(\zeta )\approx B_{0}\zeta /s$ and a locally
homogeneous electric field, $E\approx \dot{s}B_{0}/c$.

Here we consider the charged particle acceleration by the inductive electric field in the vicinity 
of the magnetic null line. Despite the simplicity of the formulation of
the problem, it is quite far from a complete solution. Even in the test
particle approximation, which describes the particle motion in the given
magnetic and electric fields, analytical solution of this problem meets serious
difficulties \cite{RecACC}. 
The reason for this is that in
the vicinity of critical points of magnetic configurations the standard
approximations adopted to describe the plasma dynamics are no longer valid.
In such regions the drift approximation, i.e., the assumption that the
adiabatic invariants are constant, can no longer be applied. In the nonadiabatic 
region the particle trajectory has the so-called ``Speiser form''. In the ultrarelativistic limit, 
the size of the nonadiabatic region and the characteristic time during which the particle moves  are 
\begin{equation}
R_{n.a.}=E/b \qquad  {\rm and} \qquad T_{n.a.}=E/bc,
\label{eq:RnaTna}
\end{equation}
respectively.

\begin{figure}
\begin{center}
\begin{tabular}{c}
\includegraphics[keepaspectratio=true,height=65mm]{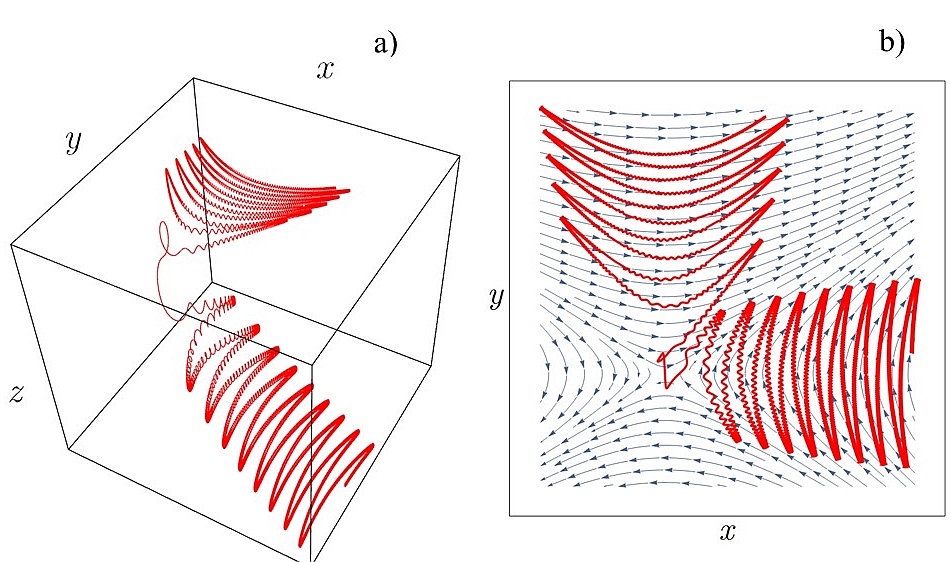}
\end{tabular}
\end{center}
\caption{ a)Trajectory of the electron accelerating along the magnetic 
X-line in the field ${\bf B}= y {\bf e}_x+b\,x {\bf e}_y$ and electric field  ${\bf E}=E {\bf e}_z$ ($b=0.5,\,E=1$). b) Trajectory projection in the $(x,y)$ plane. 
\label{Fig6}}
\end{figure}

The particle spends only a finite time interval in the nonadiabatic
region, since its motion is unstable there. After a finite time interval it
gets out of the nonadiabatic region, and gets into the drift region as it is
seen in Fig. \ref{Fig6}, where we present the trajectory of the particle moving in the 
vicinity of the magnetic field X-line with the electric field parallel to it: 
${\bf B}= by {\bf e}_x+bx {\bf e}_y$, 
${\bf E}=E {\bf e}_z$. Inside the nonadiabatic region 
the dependence of the $x$ and $y$ coordinates on time in the ultrarelativistic 
limit is given by 
\begin{equation}
x(t)=x_0 I_0(2\sqrt{bct/E}), \qquad y(t)=y_0 J_0(2\sqrt{bct/E}),
\label{eq:BreakUP-B2}
\end{equation}
where $I_0(z)$ and $J_0(z)$ are the Bessel functions and $x_0$ and $y_0$ are the particle initial coordinates.

Matching the solutions corresponding to the particle
trajectories in different regions, we can describe the particle motion and
hence the energy spectrum of particles accelerated near critical points of the magnetic configurations.
In the case of constant gradient magnetic field with $b=$constant 
the energy spectrum of accelerated particles has an exponential form. In the time dependent electromagnetic configuration 
the energy spectrum can have a power law form.

\subsection{Radiation Friction Effects on Charged Particle Acceleration}

When the particle energy exceed the level at which the radiation losses become significant the acceleration 
is less efficient. In particular, in the laser-matter interaction the radiation friction cannot be neglected for the 
laser light intensity above $10^{23}$W/cm$^2$. In the case of space plasmas, 
the radiation losses during the charged particle acceleration in the magnetic reconnection processes are
caused by backward Compton scattering and synchrotron radiation \cite{BKORF, UZD2}. A
characteristic time of the synchrotron losses for the electron with energy 
$\mathcal{E}_e$ is given by the expression 
\begin{equation}
\tau _{B}=\frac{3m_{e}^{4}c^{7}}{2e^{4}B^{2}\mathcal{E}_e}.
\end{equation}
As it was shown in Ref. \cite{BKORF}, during solar flares this effect limits
the ultrarelativistic electron energy to a value of about several tens of
GeV. 

Ultrarelativistic electron acceleration during the magnetic field line reconnection under the 
conditions of strong radiation cooling has been invoked as a model of the gamma-ray flares in the 
Crab Nebula \cite{CERUTTI}. The gamma-ray telescopes reported the detections of very bright 
high energy gamma rays ($>100$ MeV) from the Crab Nebula \cite{CRAB}, 
which suggests that the electron-positron 
pairs are accelerated to PeV ($10^{15}$eV) energies within a few days. 
For the accelerating electrons experiencing the radiation losses there 
is a balance between the electric field force 
$eE$ and the radiation reaction force due to synchrotron losses, $(2/3)r_e^2\gamma^2B^2$, {where 
 $r_e=e^2/m_e c^2=2.8\times 10^{-13}$cm is the classical electron radius.} It yields for the electron 
gamma factor 
\begin{equation}
\gamma_{rad}=\left(\frac{3eE}{2r_e^2B^2}\right)^{1/2}=\left(\frac{3E}{2 B}\frac{B_{cr}}{B}\right)^{1/2}
\end{equation}
with $B_{cr}=m_e^2c^4/e^3=6.48\times 10^{15}\,$G being the critical magnetic field of classical electrodynamics.
The emitted synchrotron photon energy is \cite{CERUTTI}
\begin{equation}
\hbar \omega_{\gamma}=\left(\frac{3\hbar}{2m_e c^2}\right)B \gamma_{rad}^2=m_ec^2\left(\frac{9 E}{4\alpha B}\right)
\approx 160\left(\frac{E}{B}\right)MeV,
\end{equation}
where $\alpha=e^2/\hbar c\approx1/137$ is the fine structure constant. 
We note that if the electrons are accelerated in the X-line vicinity
where the electric field is finite and the magnetic field vanishes, the radiation friction constraint can be mitigated.

In the adiabatic region, for $x\gg E/b$, the electron undergoes the ${\bf E}\times {\bf B}$ and gradient drift. Its 
coordinate increases with time as $x=\sqrt{c Et/b}$. Here we took into account 
the magnetic field inhomogeneity with $|B|=bx$. In the adiabatic region, the transverse adiabatic invariant 
$\mu_{\perp}=p_{\perp}^2/B$ is conserved, which results in the growth of the electron energy: 
${\cal E}_e={\cal E}_e^{(0)}(t/t^{(0)})^{1/4}$, i.e. the acceleration rate is
\begin{equation}
\dot{\cal E}_e^{(+)}=\frac{{\cal E}_e}{4t}.
\end{equation}
Here ${\cal E}_e^{(0)}$ is the electron energy with which it enters the adiabatic region, and it leaves
 the adiabatic region at time $t^{(0)}$. 
Since the size of non-adiabatic region is of the order of $
R_{n.a.}=E/b$ we can estimate these energy and time 
as ${\cal E}_e^{(0)}=e E R_{n.a.}=eE^2/b$ and $t^{(0)}=T_{n.a.}=E/bc$ (see Eq.(\ref{eq:RnaTna})), respectively. 
Radiation friction leads to the energy loss with the rate
\begin{equation}
\dot{\cal E}_e^{(-)}=-\left(\frac{{\cal E}_e}{m_ec^2} \right)^2\frac{2e^4 bEt}{3m_e^3c^4}.
\end{equation}
Using these relationships we can find the maximum electron energy, which is equal to 
\begin{equation}
{\cal E}_{e,max}=m_ec^2\left(\frac{3e^4 E^{13}}{8 b^7 (m_ec^2)^5}\right)^{1/9}.
\label{eq:EmaxRad}
\end{equation}

For the Crab nebulae (e.g. see \cite{CrNe}) this gives 
 $\gamma_{e,max}\approx10^{10}(E/B)^{13/9}$.
 
\section{ Radiation Friction and Quantum Mechanics Effects} 

\subsection{Four Regimes of Strong Electromagnetic Field Interaction with Matter}

In the limit of extremely high intensity of electromagnetic field,
the radiation friction effects begin to dominate the charged particle dynamics  \cite{MTB}. 
The electron dynamics becomes dissipative with
fast conversion of the electromagnetic wave energy into hard electromagnetic radiation, which is in the gamma-ray range for
typical laser parameters.
For laser radiation with $1\,\mu$m wavelength the radiation friction force changes the scenario
of the electromagnetic wave interaction with matter at the intensity of about $I_R\approx 10^{23}$W/cm$^{2}$. 

The probabilities of the processes involving extremely high intensity electromagnetic field interaction with 
electrons, positrons and photons are determined by several dimensionless parameters.

When the normalized dimensionless electromagnetic wave amplitude $a$ exceeds unity,
 $a$, the energy of the electron quivering in the field of the wave becomes relativistic.
Here $\lambda=2 \pi c/\omega$ with $\omega$ being the electromagnetic wave frequency.

The power emitted by an electron is proportional to the fourth power of its energy, $m_e c^2 \gamma$, \cite{LL-TP} $P_{\gamma}\approx\varepsilon_{rad} m_e c^2 \omega \gamma_e^4$.
The dimensionless parameter, 
\begin{equation}
\varepsilon_{rad}={4 \pi r_e}/{3 \lambda}=1.17\times 10^{-8} \left({1 \mu {\rm m}}/{\lambda}\right),
\label{eq:eprad}
\end{equation}
proportional to the ratio of the classical electron radius $r_e$
and the electromagnetic wave wavelength $\lambda$ characterizes the role of radiation losses. 
The maximal rate at which an electron can acquire
the energy from the electromagnetic field is approximately equal to $m_e c^2 \omega a$.
The condition of the balance between the acquired and
lost energy for the electron Lorentz factor equal to $\gamma_e=a$ shows that the
radiation effects become dominant at $a_0>a_{rad}=\varepsilon_{rad}^{-1/3}$.

QED effects become important, when
the energy of the photon generated by Thomson (Compton) scattering is of
the order of the electron energy, i.e. $\hbar \omega_m \approx m_e c^2 \gamma_e$. 
If $\gamma_e=a_0$ this yields the quantum electrodynamics 
limit on the electromagnetic field amplitude, $a_0^2/a_S>1$. Here 
the dimensionless parameter
\begin{equation}
a_S=\frac{e E_S \lambda}{2 \pi m_e c^2}=\frac{m_e c^2}{\hbar \omega}=\frac{\lambda}{\lambda_C}=4.2\times 10^{5}\left(\frac{\lambda}{1 \mu {\rm m}}\right)
\label{eq:aS}
\end{equation}
is the normalized critical electric field of quantum electrodynamics \cite{BLP}, $E_S=m_e^2 c^3/e \hbar$,
with $\lambda_C=2 \pi \hbar /m_e c=2.42\times 10^{-10}$cm 
being the Compton wavelength. 

The above obtained quantum electrodynamics limit, $a_0^2/a_S>1$, corresponds to the condition $\chi_e >1$, 
where the relativistic and gauge invariant parameter $\chi_e$,
\begin{equation}
\chi_e=\frac{\sqrt{\left(F^{\mu \nu} p_{\nu}\right)^2}}{E_S m_e c },
\label{eq:chie}
\end{equation}
{where the 4-tensor of the electromagnetic field is defined as $F_{\mu \nu}=\partial_{\mu} A_{\nu}-\partial_{\nu} A_{\mu}$ and $\mu=0,1,2,3$.  On the order of magnitude the parameter $\chi_e$ is equal to the ratio of the electric field to the critical electric field of quantum electrodynamics, $E_S$, in the electron rest frame. In particular,  it
characterizes the probability of the gamma-photon emission 
by the electron with 4-momentum $p_{\nu}$ in the field of the electromagnetic wave. }

\begin{figure}
\begin{center}
\includegraphics[keepaspectratio=true,height=65mm]{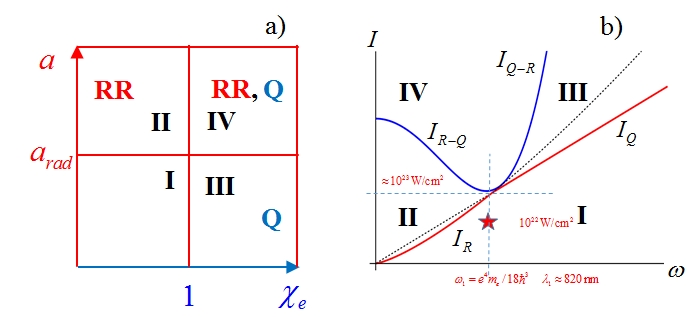}
\end{center}
\caption{ a) Regimes of electromagnetic field interaction with matter on the plane of parameters 
$a=eE/m\omega c$ and $\chi_e\approx (a/a_S)({\cal E}_e/m_ec^2)$. 
b) Curves $I_R(\omega)$, $I_Q(\omega)$ and $I_{R-Q}(\omega)$ , $I_{Q-R}(\omega)$ 
    subdivide the $(I, \omega)$ plane to 4 domains corresponding to the frame a):
(I) Relativistic electron - EM field interaction with neither radiation friction nor QED effects;
(II) Electron - EM wave interaction is dominated by radiation friction; 
(III) QED effects important with insignificant radiation friction effects;
(IV) Both QED and radiation friction determine radiating charged particle  dynamics in EM field.
The star shows the intensity achieved in experiments \cite{Ogura}.}
\label{Fig9}
\end{figure}

Using these two dimensionless parameters, $a$ and $\chi_e$, we can subdivide the $(a,\chi_e)$ plane into four 
domains shown in Fig. \ref{Fig9} a) (see Ref. \cite{SVB-2015}). If the EM field amplitude $a$ is less that $a_{rad}$ and the parameter 
$\chi_e$ is small, neither radiation friction nor QED effects are significant. For $a>a_{rad}$ and $\chi_e \ll 1$ 
the electron - EM wave interaction is dominated by radiation friction with insignificant role of QED effects. 
In the case $a<a_{rad}$ and $\chi_e > 1$ the QED effects are important with insignificant radiation friction. 
Both the QED and radiation friction determine radiating charged particle  dynamics in EM field in the limit 
$a>a_{rad}$ and $\chi_e \gg 1$.
Fig. \ref{Fig9} b) shows corresponding $(I, \omega)$ plane with 4 domains.
The curves $I_R(\omega)$ and $I_Q(\omega)$ intersect each other at the frequency equal to 
$\omega_1=e^4 m_e/18 \hbar^3$, for which the radiation wavelength is of the order of $820$nm. 
The intensity at the intersection point is about $ 10^{23}$W/cm$^{2}$.

\subsection{QED: Solution of Dirac Equation near Magnetic Null Surface}

Consistent implication of the QED effects to the theory of charged particle acceleration during the reconnection 
of magnetic field lines implies at first a thorough analysis of particle motion in an inhomogeneous electromagnetic field.
Dirac equation for the electron {4-component wave function $\psi$} in an external electromagnetic field, given by the 4-potential $A_{\mu}=(\Phi, {\bf A})$, 
is \cite{BLP}
\begin{equation}
[\gamma^{\mu}(\hat p_{\mu}-e A_{\mu})-m_e]\psi=0,
\end{equation}
{where $\gamma^{\mu}$ is a $4\times 4$ matrices (the Dirac matrices) and $\hat p_{\mu}$ is the 4-momentum operator} with $\hbar =c=1$. Near the magnetic null surface the 4-potential is equal to $A_{\mu}=(0,0,0,-Et+bx^2/2)$.

If the electric field vanishes, $E=0$, the Dirac equation for the wave function $\psi(x,y,z,t)=\varphi(x)\exp\left(-i{\cal E}t+ip_y y+ip_z z\right)$
takes the form
\begin{equation}
\left[-\frac{d^2}{dx^2}+\left(p_z-\frac{b}{2}x^2\right)^2-\sigma b x\right]\varphi=\left({\cal E}^2-m_e^2-p_y^2\right)\varphi
\label{eq:unharm-osc}
\end{equation}
with the spin $\sigma=\pm 1/2$. This is an equation for the anharmonic quantum oscillator.

For large positive $z$ component of the momentum the $\phi$ function is localized near local minima of the potential at
\begin{equation}
x_{\pm}\approx \pm\left(2 p_z/b \right)^{1/2}\left[1\mp\frac{1}{4} \left(b/2p_z^3 \right)^{1/4}\right]
\end{equation}
The electron behavior is described by the Landau theory with the magnetic field equal to $B=bx_{\pm}$.
The energy is quantized with the levels given by equation 
\begin{equation}
{\cal E}_n=\left[m_e^2+p_y^2+(2p_z b)^{1/2}(2n+1-\sigma) \right]^{1/2}, \quad n=0,1,2, ...\,.
\end{equation}

Near the magnetic null-surface for $x\to 0$ Eq. (\ref{eq:unharm-osc}) can be reduced to 
\begin{equation}
\left[-\frac{d^2}{dx^2}+b p_z x^2 -\sigma b x\right]\varphi=\left({\cal E}^2-m_e^2-p_y^2\right)\varphi.
\label{eq:unharm-osc-null}
\end{equation}
In this case, the energy is quantized with the levels 
\begin{equation}
{\cal E}_n=\left[m_e^2+p_y^2+p_z^2+p_z b(2n+1-\sigma) \right]^{1/2}, \quad n=0,1,2, ...\,.
\label{eq:EnX}
\end{equation}

In nonrelativistic limit described by the Schrodinger equation the quantum mechanics effects in the electron motion near 
the magnetic null-surface were analyzed in Ref. \cite{BLS}.

In contrast to the classical Landau results on the electron motion in the homogeneous magnetic field, 
where the characteristic electron energy $\hbar \omega_{Be}$ with $\omega_{Be}=eB/m_ec$ does not depend on the electron momentum, in the case of the 
magnetic null-surface corresponding to Eq. (\ref{eq:EnX}), the characteristic energy depends on the $z$-component 
of the electron momentum as $\hbar \omega_X$ with $\omega_X=\sqrt{ebp_z/c}$.

For non-vanishing electric field, $E \neq 0$, we seek the solution of the Dirac equation in the quasi-classical approximation, i.e. neglecting the spin effects, representing the wave function as 
\begin{equation}
\psi(x,y,z,t)=\sqrt{H(x,t)}\exp\left(-i\Theta(x,t)+ip_y y+ip_z z\right).
\end{equation}
This yields the equations for the wave function amplitude and phase 
\begin{equation}
\partial_t(H\partial_t \Theta)-\partial_x(H\partial_x \Theta)=0,
\label{eq:H-S}
\end{equation}
\begin{equation}
(\partial_t \Theta)^2-(\partial_x \Theta)^2-(Et+bx^2/2)^2=0.
\label{eq:SH-J}
\end{equation}
The solution for the amplitude equation (\ref{eq:H-S}) can be expressed via the first integrals, which  are found 
by solving the characteristic equations
\begin{equation}
\frac{dt}{\partial_t \Theta}=-\frac{dx}{\partial_x \Theta}=\frac{(\partial_{tt} \Theta-\partial_{xx} \Theta)}{\partial_t \Theta}\frac{dH}{H}
\label{eq:H-char}
\end{equation}
for known phase $\Theta(x,t)$. The equation (\ref{eq:SH-J}) for the phase has the form of the Hamilton-Jacoby equation 
of the classical relativistic particle moving near magnetic null-surface, whose motion we have analyzed above. 
Using a smallness of the magnetic field gradient we can find in the limit $t\to \infty$ that the phase depends on time and coordinate as
\begin{equation}
\Theta\approx \frac{1}{2}[(Et)^2+\sqrt{Ebt}x^2].
\label{eq:S-tx}
\end{equation}

\begin{figure}
\begin{center}
\begin{tabular}{c}
\includegraphics[keepaspectratio=true,height=65mm]{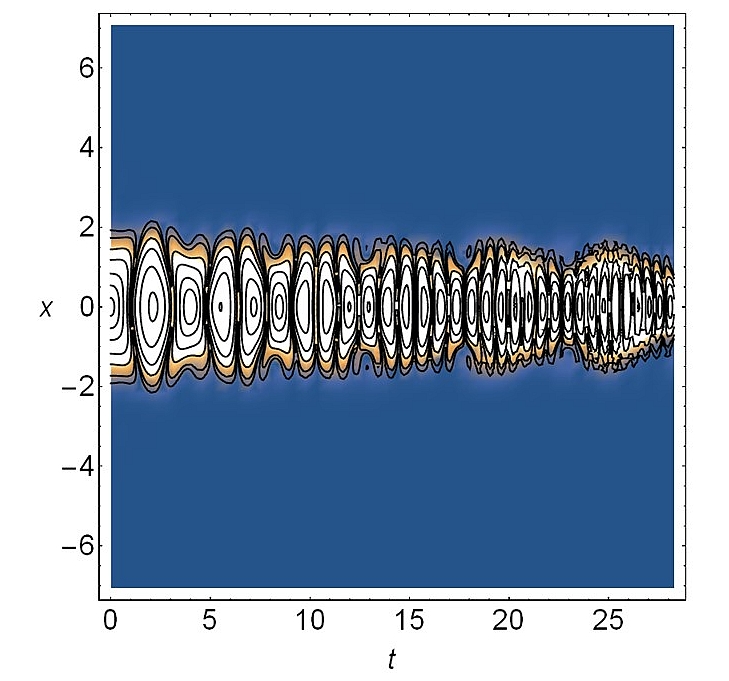}
\end{tabular}
\end{center}
\caption{Absolute value of the wave function $\phi(x,t)$ in the plane $(t,x)$ for initial condition $\phi(x,0)=\exp(-x^2)$ and $E=0.2$, $b=2$ and $m_e^2+p_y^2+p_z^2=1$ .
\label{Fig10}}
\end{figure}

In the approximation, where we neglect the spin effects, the psi-function $\psi(x,y,z,t)=\phi(x,t)\exp\left(ip_y y+ip_z z\right)$ obeys the Klein-Gordon equation
\begin{equation}
\partial_{tt} \phi-\partial_{xx} \phi+[m_e^2+p_y^2+p_z^2+(Et+bx^2/2)^2] \phi=0.
\label{eq:K-G}
\end{equation}
Its numerical solution for initial condition $\phi(x,0)=\exp(-x^2)$ and $E=0.2$, $b=2$ and $m_e^2+p_y^2+p_z^2=1$ 
is shown in Fig. \ref{Fig10}.

As we see, the local frequency of the wave-function oscillations grows with time in accordance with Eq. (\ref{eq:S-tx}).

{Obtained above expressions for the electron $\psi$-function will be used in developing in the future a rigorous theory of the radiation the charged particle  moving in the vicinity of the magnetic null line. Below we discuss the QED effects in charged particle radiation within the framework of simple theoretical model, which allows taking into account the mitigation of the radiation losses. }

\subsection{QED effects in charged particle radiation}

According to quantum electrodynamics, the electron in the strong EM field can not emit a photon with the energy exceeding the electron initial energy \cite{BLP, RITUS}. 
In the case of an electron, accelerated in the vicinity of the null line or point 
of the magnetic field and then moving in the drift region, this electron undergoes synchrotron losses. In the limit when the QED effects come into play both the radiation intensity and characteristic 
frequency of the emitted photons become lower than those in the classical case. The threshold of QED effects is determined by the dimensionless parameter $\chi_e$ defined by Eq. (\ref{eq:chie}). 
 For an electron moving in the magnetic field $B$, it is equal to $\chi_e\approx (B/B_S)({\cal E}_e/m_ec^2)$, where $B_S=m_e^2c^3/e\hbar$ is the QED critical magnetic field. The energy of the emitted photons is 
\begin{equation}
\hbar \omega_{\gamma}=\frac{{\cal E}_e\chi_e}{2/3+\chi_e}.
\label{eq:omQED}
\end{equation}
In the limit $\chi_e\ll 1$ the frequency $\omega_{\gamma}$ is equal to $(3/2)\omega_{Be}({\cal E}_e/m_ec^2)^2$ in accordance with the results of classical electrodynamics. If $\chi_e\gg 1$ the photon energy is equal to the energy of radiating electron: $\hbar \omega_{\gamma}={\cal E}_e$.

The QED effects can be incorporated into the equations of the electron motion by using the form-factor $G_e(\chi_e)$ (see Ref. \cite{QED}),
which is equal to the ratio of full radiation intensity to the intensity of the radiation emitted by classical electron. It reads 
\begin{equation}
G_e(\chi_e)=
\frac{3}{4}\int^{\infty}_0\left[\frac{4+5\chi_ex^{3/2}+4\chi_e^2x^{3}}{\left(1+\chi_ex^{3/2}\right)^4} \right]
\Phi^{\prime}(x)xdx,
\label{eq:GeChi}
\end{equation}
where $\Phi(x)$ is the Airy function.  The radiation friction force with the form-factor $G_e(\chi_e)$ 
mitigating the radiation losses can be written as  \cite{SVB-2015, QED}
\begin{equation}
{\bf f}_{rad}=-\frac{2 G_e(\chi_e)}{3c }
r_e^2 {\bf v}\left(\frac{{\cal E}_e}{m_ec^2} \right)^2\left[ \left({\bf E}+\frac{1}{c}{\bf v}\times {\bf B}\right)^2
-\left(\frac{1}{c}{\bf v}\cdot {\bf E}\right)^2\right].
\label{eq:FradGeChi}
\end{equation}
\begin{figure}[tbp]
\includegraphics[keepaspectratio=true,height=65mm]{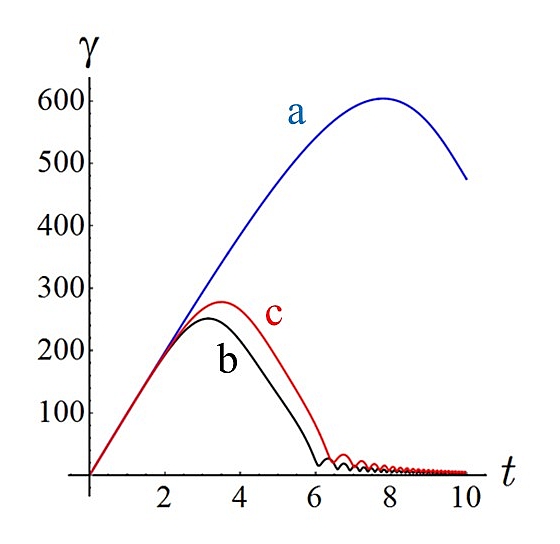}
\caption{Normalized electron energy vs time for the electron accelerated in the vicinity of the magnetic $X$-line with normalized magnetic field gradient $b=100$ in the electric field $E=100$. 
a) Regime I: $\varepsilon_{rad}=0$, $\chi_e=0$. b) Regime II: $\varepsilon_{rad}=10^{-6}$, $\chi_e=0$.  
c) Regime IV: $\varepsilon_{rad}=10^{-6}$, $b_S=5.1\times 10^{3}$.
}
\label{Fig11}
\end{figure}

In Fig. \ref{Fig11}, we present the time dependence of normalized energy $\gamma={\cal E}_e/m_ec^2$
 for the electron accelerated in the vicinity of the magnetic $X$-line
 with the electric field parallel to it: ${\bf B}= by {\bf e}_x+bx {\bf e}_y$, 
${\bf E}=E {\bf e}_z$. We use dimensionless electric field $eET_{n.a.}/m_e c=eE^2/m_ebc^2=100$ (for $T_{n.a.}$ see Eq. (\ref{eq:RnaTna})). The parameter $\varepsilon_{rad}$ characterizing the radiation friction 
is equal to $\varepsilon_{rad}=r_e/R_{n.a.}=e^2b/m_eEc^2$. The normalized QED field $B_S$ is $b_S=eB_ST_{n.a.}/m_ec=Em_ec/b\hbar$. The parameters correspond to the cases: a) when there are no radiation friction and quantum effects (i.e. with $\varepsilon_{rad}=0$, $\chi_e=0$, it is in the regime I in Fig. \ref{Fig9}); b) when there is radiation friction but no quantum effects, $\varepsilon_{rad}=10^{-6}$, $\chi_e=0$,
 it is in the regime II in Fig. \ref{Fig9}); c) when there are both the radiation friction and quantum effects, $\varepsilon_{rad}=10^{-6}$ and $b_S=5.1\times 10^{3}$, it is in the regime IV in Fig. \ref{Fig9}). As we see,
the maximum energy is achieved in the case (a) when there are no radiation friction and quantum mechanics effects. The radiation friction limits the maximum energy in the case (b), 
when there is radiation friction but no quantum effects in accordance with Eq. (\ref{eq:EmaxRad}). The quantum mechanics effects mitigating the radiation losses increase the electron energy in the case (c), when both the radiation friction and quantum effects 
are taken into account.

\section{Conclusion}

In conclusion, what we will learn about the reconnection of the magnetic field lines, {particularly,} with the relativistic laser plasmas 
is the knowledge on the reconnection in the range of the regimes from the MHD to QED. The experimental and theoretical 
studies of the reconnection in the plasma irradiated by extremely high power laser radiation will provide an opportunity to
reveal novel properties of the magnetic reconnection in nontrivial magnetic topology, of reconnection of generalized 
vorticity in collisionless plasmas, of obtaining novel knowledge on relativistic regimes with strong electric field generation
due to the magnetic field annihilation, and of the role of the radiation friction and QED effects on the charged particle acceleration 
and radiation during the magnetic reconnection.

\section*{Acknowledgments}

{My thanks to Plasma Physics Division of the European Physical Society for the Alfv\'en prize. 
I am indebted to the physicists who are not longer with us: Sergei Ivanovich Syrovatskii, my graduate advisor,
introduced me to the theory of magnetic field line reconnection; Vitaly Lasarevich Ginzburg instilled in me a love for 
astrophysics; Gurgen Ashotovich Askar'yan, who was full of outstanding ideas; Ferdinand Cap with whom I used to work on charged particle acceleration.  I appreciate
 discussions of various aspects of magnetic
reconnection with G. Bertin, F. Califano, B. Coppi, G. I. Dudnikova, E. Yu. Echkina, T.
Zh. Esirkepov, D. Farina, Y. Gu, I. N. Inovenkov, M. Kando,
P. Kaw, O. Klimo, J. Koga, K. Kondo, G. Korn, G. R. Kumar, E. Lazzaro, T. V. Liseikina, L. Nocera, M. A. Olshanetskij, F. Pegoraro, V. V. Pichushkin, F. Porcelli, R. Pozzoli, J.-I.
Sakai, A. S. Sakharov, P. V. Sasorov, K. Schindler, A. Sen, T. Tajima, G. E. Vekstein,
S. Weber, A. Yogo, and L. M. Zelenyi. This work was partially funded by ImPACT Program of Council for Science, Technology
and Innovation (Cabinet Office, Government of Japan).}


\end{document}